\documentstyle[prd,aps]{revtex}
\begin{document}
\begin{titlepage}
\title{CMB anisotropy predictions for a model of double inflation}
\author{Julien Lesgourgues$^{1}$\ and D. Polarski$^{1,2}$\\
\hfill \\
$^1$~{\it Lab. de Math\'ematiques et de Physique Th\'eorique, UPRESA 6083 
CNRS}\\
{\it Universit\'e de Tours, Parc de Grandmont, F-37200 Tours (France)}\\
\hfill\\
$^2$~{\it D\'epartement d'Astrophysique Relativiste et de Cosmologie},\\
{\it Observatoire de Paris-Meudon, 92195 Meudon cedex (France)}\\
\hfill \\}

\date{22 May 1997}
\maketitle

\begin{abstract}
We consider a double-inflationary model with two massive scalar fields 
interacting only gravitationally in the context of a flat cold 
dark matter (CDM) Universe. The cosmic microwave background (CMB) temperature 
anisotropies produced in this theory are investigated in great details for a 
window of parameters where the density fluctuations power spectrum $P(k)$\ is 
in good agreement with observations. The first Doppler (``acoustic'') peak 
is a crucial test for this model as well as for other models. 
For values of the cosmological parameters close to those of standard CDM, our 
model is excluded if the height of the Doppler peak is sensibly higher than 
about three times the Sachs-Wolfe plateau. 
\end{abstract}

\end{titlepage}

\section{Introduction}
The inflationary paradigm \cite{lindekolb} has been an exciting theoretical 
development in our understanding of some basic open issues in Big Bang 
cosmology. Indeed, it provides 
an elegant explanation to the horizon problem as well as to the flatness 
problem. In addition, it incorporates in a natural way the creation of 
primordial fluctuations \cite{hawking} which will eventually grow through 
gravitational instability into the large scale structures as we observe 
them today in the Universe on cosmological scales. 
It is then possible to calculate the fluctuation spectra with great accuracy 
for all models, 
using possibly numerical methods \cite{salopek}. One still has to combine 
these primordial spectra with some assumptions about the Universe and its 
matter content.
The CDM model \cite{davis} has been 
extensively studied as the simplicity of its basic assumptions is very 
appealing. However, even before the first COBE DMR data there was 
observational evidence from the APM survey \cite{maddox} showing that 
standard CDM does 
not agree with observations. When correctly normalized on very large 
cosmological scales using the COBE DMR data of the CMB anisotropies on 
large angular scales, standard CDM has too much power on small scales.

Double Inflation \cite{lev}, \cite{NPB92} is an attempt to reconcile the 
CDM model with observations in the following way: instead of having a purely 
scale-invariant 
(Harrison-Zel'dovich) spectrum of adiabatic fluctuations one has now a 
spectrum possessing a characteristic scale. As this spectrum rests on a 
well-defined inflationary model, 
in other words there is an underlying effective 
Lagrangian which defines completely the dynamics of the inflationary 
background and the properties of the various perturbations spectra,
it is possible to study in detail, with numerical methods when 
necessary, the 
fluctuations spectra generated in this model. This was done already for both 
density (scalar) fluctuations \cite{PRD94a},~\cite{PRD94b} and tensorial 
fluctuations or gravitational waves (GW's) \cite{PLB95}.
Another otherwise attractive attempt to reconcile CDM with the observations in 
the same spirit, is to replace the primordial perturbation spectrum by a 
tilted one, i.e., one for which $n<1$. Such models were considered in 
particular after release of the first COBE DMR results which showed that 
standard CDM would give too much power on small scales, for example 
$\sigma_8\sim 1.3$. While a tilted spectrum does render 
CDM more compatible with observations, it is not in accordance with all the 
observations on large and small scales though a value $n\approx 0.8$ comes 
closest to it \cite{cen}. 
The CDM paradigm is appealing enough so that also tilted CDM models with some 
change of the cosmological parameters have been considered afterwards. An 
interesting 
possibility is a higher baryon fraction in the Universe \cite{liddle} but it 
turns out that, in order to have a value for $\sigma_8 \approx (0.6-0.7)$, 
one would need $\Omega_b\sim 15\%$ and, though not excluded, there is no 
compelling evidence for such a high baryon fraction.  
Of course, there are still other ways to depart from the standard CDM 
paradigm, another interesting possibility being a change of the matter 
content of the Universe, thereby changing the power spectrum as it is seen 
today (see, 
for example, \cite{Lyth} and references therein). This can be achieved 
if one considers a mixture of cold dark matter with a certain amount of hot 
dark matter. Yet another possibility is to consider 
universes with open spatial geometries as one would have in any case once 
there is compelling evidence for $\Omega<1$. 
\par 
So double inflation is an attempt to cure the problem by a change in the 
primordial spectrum only, leaving the other parameters otherwise unchanged, 
or in any case close to their ``canonical'' values.
The purpose of this work is to extend the study of double inflation to the 
CMB fluctuations up to small angular scales. With the advent of the next 
generation experiments whose aim is a very high precision measurement of the 
CMB anisotropies up to $l\approx 1500$, like, for example, 
the satellite mission {\it PLANCK Surveyor}, it is clear 
that the constraints coming from those observations will be crucial regarding 
the viability of the various existing models. 

The outline of this work is as follows. In Sec. \ref{sec.2}, we review 
for completeness the model we consider 
here and some of its peculiarities. In Sec. \ref{sec.3}, 
we find the window of 
allowed parameters after constraining the density power spectrum $P(k)$   
and compare our model with other ones. In Sec. \ref{sec.4}, we find the CMB 
anisotropies for the selected window of parameters. Finally in Sec. 
\ref{sec.5} we give a summary and short discussion.
 
\section{The double-inflationary background and the fluctuations}\label{sec.2}
We give here a short description of our model of double inflation,
starting with the homogeneous background.
We consider the following Lagrangian density describing matter and
gravity
\begin{equation}
L=-{R\over {16\pi G}}+{1\over 2}(\phi_{h,\mu}\phi_h^{,\mu}-m_h^2\phi_h^2) 
+{1\over 2}(\phi_{l,\mu}\phi_l^{,\mu}-m_l^2 \phi_l^2)~,
\end{equation}
where $\mu=0,..,3, c=\hbar=1$. The background space-time metric has the form
\begin{equation}
ds^2=dt^2-a^2(t)\delta_{ij}dx^i dx^j~, \qquad\ i,j=1,2,3.
\end{equation}
The first period of inflation is driven by the
heavy scalar field and we will see that the interesting part of the spectrum 
corresponds to the end of this first inflation. The homogeneous background 
is treated classically, it is determined by the scale factor $a(t)$ and the 
two scalar fields $\phi_h,\phi_l$. 
\par
A crucial ingredient of any inflationary model is the generation of 
inhomogeneous perturbations, of quantum mechanical origin, superimposed on 
the homogeneous background. 
For their description we consider a perturbed FRW background whose metric, 
in the longitudinal gauge, reads
\begin{equation}
ds^2=(1+2\Phi)dt^2-a^2(t)(1-2\Psi)\delta_{ij}dx^idx^j~,
\end{equation}
(in Bardeen's notation~\cite{bardeen}, $\Phi=\Phi_A, \Psi=-\Phi_H$).
We are interested in the spectrum of
growing adiabatic perturbations which arise from the
vacuum fluctuations of the scalar fields $\phi_h$ and $\phi_l$. They are 
Gaussian and the power spectrum $\Phi^2(k)$ of the gravitational potential, 
defined through 
\begin{equation}
\langle \Phi_k \Phi^*_{k'}\rangle=\Phi^2(k)~\delta ({\bf k}-{\bf k'})~, 
\end{equation}
characterizes them completely.
Some of the features of the spectrum when the intermediate matter-dominated 
stage is pronounced are relevant also to cases where it is not.
For scales crossing the Hubble radius when both scalar fields are
in the slow rolling regime, the spectrum of growing adiabatic
perturbations, when those scales are outside the Hubble radius during the
matter-dominated stage (assuming $a(t)\propto t^{\frac{2}{3}}$ at the
present time), is given by~\cite{NPB92}
\begin{eqnarray}
k^{3\over 2}\Phi(k) 
& \simeq & {{\sqrt{24\pi G m_h^2}}\over 5}\sqrt{s\ln {{k_f}\over k}}
\qquad\ k\ll k_f~,\label{eq:log}
\end{eqnarray}
where the rhs has to be taken at $t=t_k$, the first Hubble radius crossing 
time and the quantity $s(t)$ is the total number of $e$-folds from time $t$ 
till the end of the second inflation.
The wave number $k_f$ corresponds to the scale crossing the Hubble radius 
near the end of the first inflation.
An important point is seen from Eq. (\ref{eq:log}), namely that the upper 
part of the spectrum, corresponding to small $k'$s or very large scales, is not
flat but has a logarithmic dependence $\propto \ln^{1\over 2}\frac{k_f}{k}$.  
This is why the naive picture of two plateaus fails and one has to calculate 
the spectrum with accuracy if the model is to be confronted with 
observations. This remains true when the intermediate matter-dominated stage 
is absent and has crucial observational consequences.

We introduce the parameter $p\equiv {{m_h}\over{m_l}}$, higher $p$ values 
correspond to bigger ``steps'' (see figure 1).
The height of the ``step'' $\Delta_k$ between a scale on
the upper ``plateau'' (still inside our visible universe today) and a scale 
at the beginning of the lower plateau, is given by
\begin{equation}
\Delta_k \simeq 0.13 p \ln^{\frac{1}{2}}\frac{k_f}{k}, \qquad\ \qquad\ k\ll
k_f~.
\end{equation}
For $p<25$, the observationally interesting values,
the evolution of the background during
the transition between the two main inflationary stages is still 
inflationary, in the sense that $\ddot{a}>0$~\cite{PRD94a}.
In order to make accurate comparison with the observations for models without 
a pronounced intermediate matter-dominated stage, physical scales 
of our spectra are defined with the help of the quantity $k_b$, the
scale where the extrapolated upper part intersects the lower plateau. 
If $\phi_l\simeq 3 M_P$ towards the end of the first 
inflation, the second inflationary stage has the right amount of expansion 
and puts the excess of power on the right scales. A tiny change in this 
initial value is enough to shift the spectrum in $k$\ space while leaving the 
form of the spectrum practically unaltered.

\begin{figure}[htbp]
\setlength{\unitlength}{0.1bp}
\begin{picture}(4139,2483)(0,0)
\put(2278,51){\makebox(0,0){$k \quad ({\mathrm h}\,\,{\mathrm Mpc}^{-1})$}}
\put(100,1341){%
\makebox(0,0)[b]{\shortstack{$k^{3/2} \Phi(k)$}}%
}
\put(3956,151){\makebox(0,0){10}}
\put(2837,151){\makebox(0,0){1}}
\put(1719,151){\makebox(0,0){0.1}}
\put(600,151){\makebox(0,0){0.01}}
\put(540,2272){\makebox(0,0)[r]{0.0001}}
\put(540,251){\makebox(0,0)[r]{1e-05}}
\end{picture}
\caption[]{\small The primordial power spectrum $k^{3/2} \Phi(k)$ 
is plotted for different values of $p$: from top to bottom on the right, 
$p=8,~10,~12,~15,~20,~25$.
All spectra have the characteristic scale $k_b=1\,h\,{\rm Mpc}^{-1}$,
and are normalized to COBE.
At small scales, the power spectra decrease quickly with growing $p$.
At large scales the effect of $p$ is opposite, but almost negligible.}
\label{fig.phi}
\end{figure}
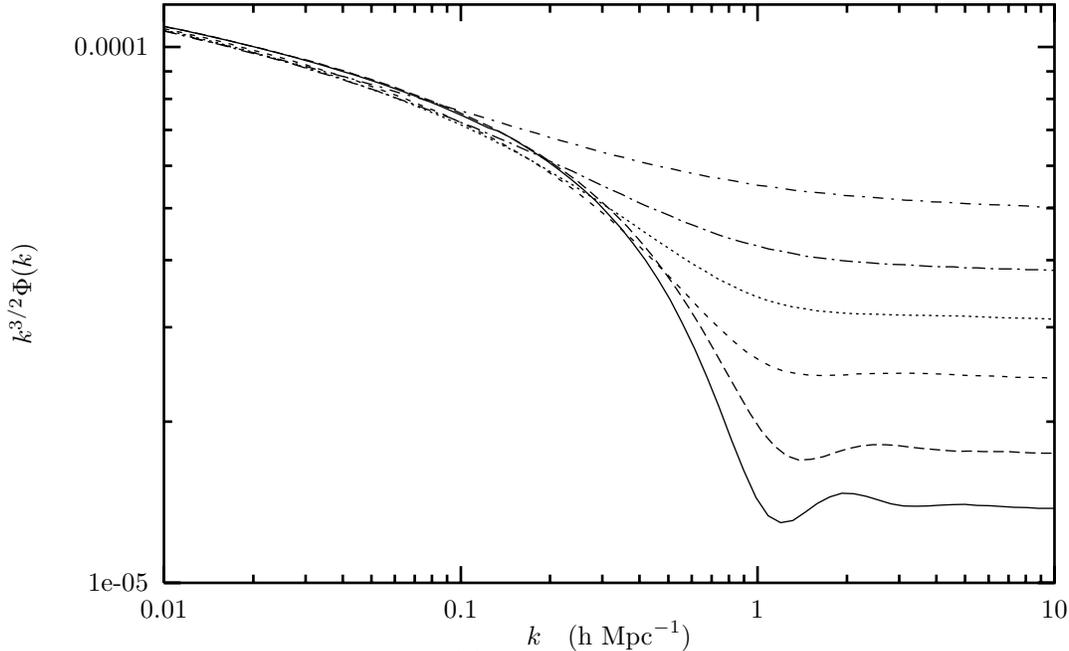

We turn now our attention to the power spectrum $P(k)$ defined through 
$\langle \delta_k \delta^*_{k'}\rangle=P(k)~\delta ({\bf k}-{\bf k'})$. 
Perturbations in the linear regime grow at different rates depending on the 
relation between their wavelengths, the Jeans length and the Hubble radius. 
The resulting amplitude for different scales is encoded in the so-called 
transfer function $T(k, t_0)$ 
\begin{equation}P(k,t_0)=\frac{4}{9}\frac{k^4}{H_0^4}~\Phi^2(k)~T^2(k),
\label{eq:transfer}\end{equation}
where, by definition, $T(k\rightarrow 0)=1$. $T(k)$ is computed numerically 
once assumptions are made about the matter content of the Universe and
other cosmological parameters like $\Omega_0$ and $h$. 
We do not use here the transfer function for the standard CDM model given 
by~\cite{bbks}
\begin{equation}
T(q)=\frac{\ln (1+2.34q)}{2.34q}\lbrack 1+3.89q+(16.1q)^2+
(5.46q)^3+(6.72q)^4\rbrack^{-\frac{1}{4}}~,
\end{equation} 
where $q\equiv \frac{k}{\Omega_0 h^2 {\rm Mpc}^{-1}}$, $\Omega_0=1$, $h=0.5$,
nor even for different shape parameters $\Gamma$. We use rather the more 
accurate transfer function that is computed numerically with the help of the 
code {\sc cmbfast} (Seljak and Zaldarriaga \cite{SZ}).
So we repeat the study of the observational constraints on $P(k)$, in order to 
find the allowed region in the ($p,~k_b$) plane, with an accuracy that matches 
the precision of the CMB anisotropies computation by Seljak's and 
Zaldarriaga's code {\sc cmbfast}. 
We assume tacitly everywhere $\Omega_0=1$, in accordance with the standard 
picture of inflation. 
\par
Another point that deserves special attention is the production of a 
primordial gravitational waves (GW's), or tensorial perturbations, background 
in this model~\cite{PLB95}.  
The tensorial perturbations $h_{ij}$ are given by 
\begin{equation}
h_{ij}=\sqrt{32\pi G}\phi e_{ij},
\end{equation}
where $\phi$ is a massless scalar field while the polarization tensor 
satisfies $e_{ij}e^{ij}=1$. Later, we will need their power spectrum $h(k)$ 
defined by
\begin{equation}
\langle h_{{\bf k},\lambda} h^*_{{\bf k'},\lambda '}\rangle \equiv h^2(k)
\delta^{(3)}({\bf k}-{\bf k'})\delta_{\lambda \lambda '}~,
\end{equation}
where $\lambda =1,2$ denotes the two polarization states.
Usually the importance of this background lies mainly in the fact that when 
its contribution to the temperature anisotropy on large angular scales is 
not too small, this balance between scalar and tensorial perturbations 
allows a smaller power spectrum $P(k)$ for given normalization. 
Though in our model the contribution of the GW 
background to the temperature anisotropy is definitely subdominant compared 
to that of the scalar adiabatic perturbations, 
it is interesting to note that the relations 
for small multipoles $l$ which hold for single-field slow-roll inflation, do 
not apply in this case. In our models we have in particular
\begin{equation}
\frac{\langle |a_{2m}|^2\rangle_{GW}}{\langle |a_{2m}|^2
\rangle_{AP}}=\frac{C_2^T}{C_2^S} \ll {\cal K}_2~|n_T|\simeq 7(1-n)\approx 1,
\label{TS}
\end{equation}  
as can be seen in figure \ref{fig.ratio}, $n_T$, resp. $n$ being the 
spectral indices of the GW, resp. the scalar perturbations 
($n\equiv 1+\frac{d\ln k^3\Phi^2(k)}{d\ln k},~ 
n_T\equiv \frac{d\ln(k^3h^2(k))} {d\ln k}$).
Relation (\ref{TS}) would become a powerfull 
discriminative test between single field inflation and our double 
inflationary model provided the GW contribution to the CMB anisotropy can be 
separated from the adiabatic scalar perturbations contribution. 

\section{Constraining the power spectrum $P(k)$}\label{sec.3}

\subsection{Power spectrum normalization}

A quite accurate normalization of the power spectrum is now possible using 
past years measurements of the CMB anisotropies on angular scales of 
a few degrees, in particular COBE DMR.
We normalize our spectra to the value of
$C_{10}$\ extrapolated from the experimental
bounds on $Q_{rms-ps|n=1}$, the quadrupole predicted for a Harrison-Zel'dovich
spectrum, since this multipole is minimally dependent on the spectral 
indices $n$ and $n_T$. A joint analysis of Tenerife Dec$=+40^\circ$\ and 
two-year COBE data gives \cite{HA} $Q_{rms-ps|n=1}=21\pm1.6 \, \mu K$, which 
means $10(10+1)C_{10}=(9.5\pm1.9)\times10^{-10}$
(the error bars take into account both
sample and cosmic variances).
The four-year COBE DMR result \cite{BE} is smaller:
$Q_{rms-ps|n=1}=18\pm1.6 \,\mu K$, i.e.,
$10(10+1)C_{10}=(6.6\pm1.2)\times10^{-10}$.
In the following, we will take $10(10+1)C_{10}=6.6\times 10^{-10}$, 
keeping in mind the possibility of a higher value.

\subsection{The window in parameter space}
\label{sec.pow.spec}

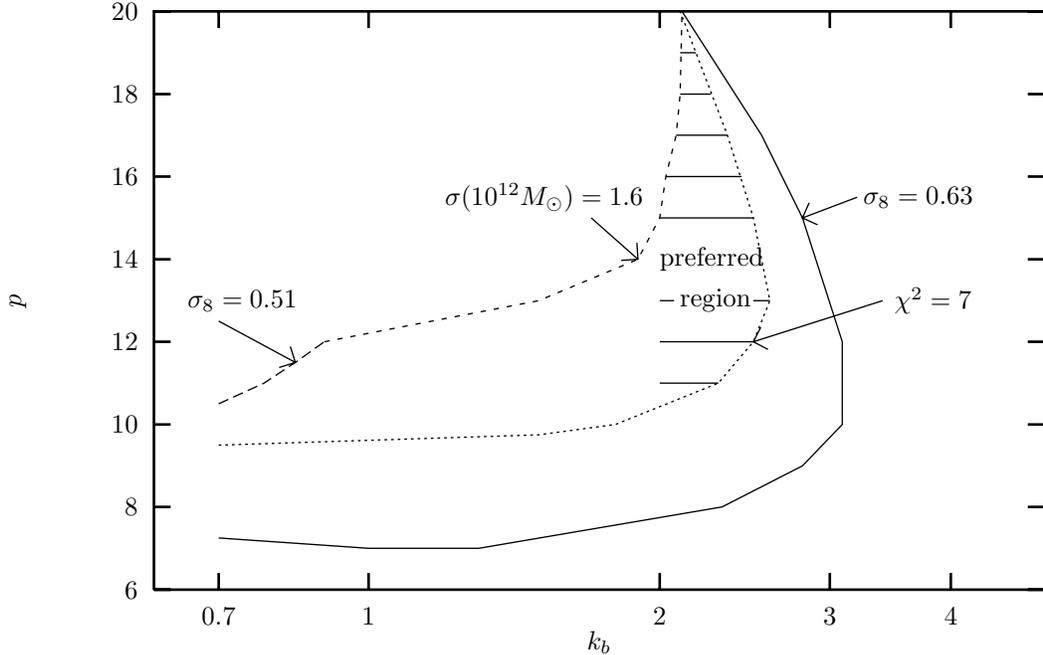
\begin{figure}[htbp]
\setlength{\unitlength}{0.1bp}
\begin{picture}(4139,2483)(0,0)
\put(2583,1341){\makebox(0,0)[l]{region}}
\put(2506,1497){\makebox(0,0)[l]{preferred}}
\put(3391,1341){\makebox(0,0)[l]{$\chi^2=7$}}
\put(1697,1731){\makebox(0,0)[l]{$\sigma(10^{12} M_\odot)=1.6$}}
\put(727,1341){\makebox(0,0)[l]{$\sigma_8=0.51$}}
\put(3274,1731){\makebox(0,0)[l]{$\sigma_8=0.63$}}
\put(2278,51){\makebox(0,0){$k_b$}}
\put(100,1341){%
\makebox(0,0)[b]{\shortstack{$p$}}%
}
\put(3603,151){\makebox(0,0){4}}
\put(3147,151){\makebox(0,0){3}}
\put(2506,151){\makebox(0,0){2}}
\put(1409,151){\makebox(0,0){1}}
\put(844,151){\makebox(0,0){0.7}}
\put(540,2432){\makebox(0,0)[r]{20}}
\put(540,2120){\makebox(0,0)[r]{18}}
\put(540,1809){\makebox(0,0)[r]{16}}
\put(540,1497){\makebox(0,0)[r]{14}}
\put(540,1186){\makebox(0,0)[r]{12}}
\put(540,874){\makebox(0,0)[r]{10}}
\put(540,563){\makebox(0,0)[r]{8}}
\put(540,251){\makebox(0,0)[r]{6}}
\end{picture}
\caption[]{\small The main constraints 
are plotted on this diagram in parameters
space. The preferred region corresponds to $10<p<20$, $2 \,h\,{\rm Mpc}^{-1} 
<k_b<2.5 \,h\,{\rm Mpc}^{-1}$.} 
\label{fig.domain}
\end{figure}

Once normalized to COBE DMR, a robust constraint on the 
matter power spectrum $P(k)$ comes from the value of $\sigma_8$\ (the
variance of the total mass fluctuation in a sphere of radius
$R=8\, h^{-1}{\rm Mpc}$), for bright galaxies this quantity is close to 
unity~\cite{peebles}. 
This is what is usually called the ``optical'' $\sigma_8$. 
White, Efstathiou, and Frenk \cite{WE} find 
\begin{equation}
\label{sigma8}
\sigma_8=0.57\pm0.06~.
\end{equation}
These bounds single out a region for our two free parameters $p$\ 
and $k_b$ as can be seen in figure \ref{fig.domain} for $h=0.5$,
$\Omega_B h^2=0.015$\ and $C_{10}=6.6\times 10^{-10}$
(small variations of $h=0.5$\ and  $\Omega_B$\ 
will be considered in Sec. \ref{sec.4}).
We get a lower bound for $p$: $p>8$ is required to obtain small enough 
power on this scale. 

An upper bound can be found using some constraints 
at even smaller scales (but still using the power spectrum obtained in the 
framework of linear theory and extrapolated to these scales),
deduced from observations of galaxies and quasars at high redshifts. 
In order to explain the formation of these
objects, one must put a lower limit on the linear
power spectrum at the corresponding scale. For instance,
estimates of the mass fraction in host galaxies of quasars at
$z=4$\ \cite{HE} require the following lower bound \cite{GM}:
\begin{equation}
\sigma(10^{11} M_\odot) \geq 2.2\pm 0.5~,
\end{equation}
where $\sigma(M)$\ stands for the variance of the total mass 
fluctuations in a sphere of mass $M$\ today, assuming linear evolution.
Similarily, since large galaxies seem to have formed as early as $z=1$,
we have the lower bound \cite{HE}:
\begin{equation}
\sigma(10^{12} M_\odot) \geq 2.0\pm 0.4~.
\end{equation}
The limit $\sigma(10^{12} M_\odot)=1.6$, which turns out to be
even more constraining for our model, is plotted in figure 
\ref{fig.domain} for our standard set of parameters and we see that in order 
to have enough power at these small scales, we must exclude any $p>20$.

Redshift surveys provide us with a huge amount of data
giving indications about the power spectrum 
at scales ranging from 15 to 300 $h^{-1}\,{\rm Mpc}$.
We compare our model with the count-in-cell analysis of large scale
clustering of the Stromolo-APM redshift survey. After normalization
of the spectra to $\sigma_8=1$, in order to get the correlation function
of optical galaxies in redshift space, we compute the variance
$\sigma_l^2$\ in cells of size $l\,\,h^{-1}{\rm Mpc}$, and
compare it with the Stromlo-APM data \cite{LE}, consisting
of nine points, assumed to be independant, with error bars treated 
as $2\sigma$ ones. 
A $\chi^2$\ analysis selects the best-fitting parameters $p$\ and 
$k_b$. Since we have seven degrees of freedom, $\chi^2\leq7$\ means 
that we are in very good agreement with the data.
Inside our previously allowed window, 
this requires $p>10$\ and 
$k_b \leq 2.5 \,h\,{\rm Mpc}^{-1}$. If we apply the same test,
for instance, to tilted models, it turns out that $\chi^2\leq7$\ implies
a too low spectral index, namely $n\leq0.6$. 

Finally, we compare our spectra with the results 
of Kolatt and Dekel \cite{KD} derived from peculiar velocities of galaxies.
Using the Mark III catalog and the reconstruction of the density field from 
{\sc POTENT}, 
a direct estimate of the power spectrum is given for three values of 
$k$, independently of the bias.
For all values of $p$ and $k_b$ inside the previously found window, the 
double-inflationary power spectrum is too low in order to agree at the 
$1\sigma$\ level with all three points. However, for the highest $k_b$'s, 
the disagreement is rather small. Indeed, for $k_b > 2\,h\,{\rm Mpc}^{-1}$, 
the spectrum agrees with all three points at the $1.5 \sigma$ level.

Finally, our preferred region is then 
$2\,h\,{\rm Mpc}^{-1} \leq k_b \leq 2.5 \,h\,{\rm Mpc}^{-1}$ and $10<p<20$, 
as can be seen in figure \ref{fig.domain}.
In other words, the position of the step in real space should be
\begin{equation} 
2.4 \, h^{-1}\,{\rm Mpc} <\frac{2\pi}{k_b}<3 \, h^{-1}\,{\rm Mpc}~,
\end{equation}
and the mass of the heavy inflaton 
\begin{equation}
m_h=(3.6 \pm 0.2)\times10^{-6} {\rm M_P}~.
\end{equation} 

Let us come back to the dependence of these results on
the COBE normalization. If the value of the quadrupole is higher 
than that given by the four-year COBE data, our results for $p$ would
be approximately the same, but the allowed window for $k_b$ would
shift to lower values, in order
to cancel the power increase on small scales. This would be an
improvement for double-inflationary models, in the sense that higher values 
of the peculiar velocities would then be reached.
For instance, if we normalize the spectra
using $Q_{rms-ps|n=1}=21\pm1.6 \, \mu K$\
\cite{HA}, we find a narrow window in which all
constraints, including peculiar velocities, are satisfied at the 
$1\sigma$\ level: $12 \leq p \leq 16$ and 
$1\,h\,{\rm Mpc}^{-1} \leq k_b \leq 1.1 \,h\,{\rm Mpc}^{-1}$. 

\subsection{Comparison with other models} 

In table \ref{table1}, we compare the results of the previous tests
for several models: our model (we choose parameters at the center of the 
previously found window, viz. $p=12, \, k_b=2.3 \,h\,{\rm Mpc}^{-1}$), 
``standard CDM'' with a flat spectrum, 
and two tilted models (with and without tensor contribution); 
all models have the same transfer function
with $h=0.5$, $\Omega_B h^2=0.015$. 
In order to make the relevant comparison, we select the values
of the spectral indices that give $\sigma_8=0.60$. We find
$n=0.70$ (without tensors) or $n=0.845$ (with tensors, assuming
$n_T=n-1$\ and $C^T_2/C^S_2=7(1-n)$).   

\begin{table}[htbp]
\begin{center}
\begin{tabular}{c| c c c c c}
& sCDM & D.I. & tilted (S)& tilted (S+T) & Obs.\\
\hline
$\sigma_8$                &1.13 & 0.60 & 0.60 & 0.60 & 0.57$\pm$0.06 \\
$\sigma(10^{11} M_\odot)$ &7.84 & 2.24 & 2.94 & 3.51 & $\geq$ 1.7    \\ 
$\sigma(10^{12} M_\odot)$ &5.72 & 1.84 & 2.31 & 2.66 & $\geq$ 1.6    \\
$\chi^2         $         &43   & 6.5  & 13   & 26   & $\leq$ 7      \\
$P(k=0.061)$              &9360 & 4200 & 3650 & 3200 & 8157$\pm$3127 \\
$P(k=0.102)$              &7590 & 2850 & 2540 & 2390 & 4620$\pm$1240 \\
$P(k=0.172)$              &4800 & 1450 & 1370 & 1400 & 1968$\pm$495  \\
\end{tabular}
\end{center}
\caption{\small 
The results of a few tests are given together with the observational bounds 
for several models: standard CDM, double-inflation with $p=12, \, k_b=2.3 
\,h\,{\rm Mpc}^{-1}$,
and tilted models wich are found to be consitent with $\sigma_8=0.60$,
i.e., $n=0.70$ (without tensors) and $n=0.845$ (with tensors).
All these results are based on $h=0.5$,
$\Omega_B h^2=0.015$\ and four-year COBE normalization. The wavenumber $k$ 
is expressed in $h\,{\rm Mpc}^{-1}$\ and $P(k)$\ in $h^{-3}{\rm Mpc}^3$.}
\label{table1}
\end{table}

It is clear from the table that our double-inflationary model gives 
better results than the chosen tilted models, namely a lower $\chi^2$\ 
and higher peculiar velocities. Of course,
most authors favor higher values of the spectral index \cite{WS}, 
but then the constraint (\ref{sigma8}) on $\sigma_8$\ is violated.
Therefore, in the framework of a flat CDM universe with its standard 
cosmological parameters
($h\simeq0.5$, $\Omega_B h^2 \simeq 0.015$), no inflationary scenario
with less than two free parameters (in addition to the overall 
normalisation) is compatible with observations while, with respect to 
constraints considered so far, double-inflation does.

\begin{figure}[htbp]
\setlength{\unitlength}{0.1bp}
\begin{picture}(4139,2483)(0,0)
\put(1439,1814){\makebox(0,0)[r]{Peacock \& Dodds}}
\put(1439,1914){\makebox(0,0)[r]{Kolatt \& Dekel}}
\put(1439,2014){\makebox(0,0)[r]{Tilted (S+T)}}
\put(1439,2114){\makebox(0,0)[r]{Tilted (S)}}
\put(1439,2214){\makebox(0,0)[r]{Standard CDM}}
\put(1439,2314){\makebox(0,0)[r]{Double Inflation}}
\put(2278,51){\makebox(0,0){$k \quad ({\mathrm h}\,\,{\mathrm Mpc}^{-1})$}}
\put(100,1341){%
\makebox(0,0)[b]{\shortstack{$P(k) \quad ({\mathrm h}^{-3}{\mathrm Mpc}^{3})$}}%
}
\put(3956,151){\makebox(0,0){1}}
\put(3117,151){\makebox(0,0){0.1}}
\put(2278,151){\makebox(0,0){0.01}}
\put(1439,151){\makebox(0,0){0.001}}
\put(600,151){\makebox(0,0){0.0001}}
\put(540,2314){\makebox(0,0)[r]{10000}}
\put(540,1283){\makebox(0,0)[r]{1000}}
\put(540,251){\makebox(0,0)[r]{100}}
\end{picture}
\caption[]{\small 
Power spectra for the previously selected models (see table
\ref{table1}) plotted together with the estimates of Peacock and Dodds 
(rescaled to $\sigma_8=0.6$) and the results of Kolatt and Dekel. All error 
bars are at $1\sigma$ level.} 
\label{fig.pow.spec}
\end{figure}
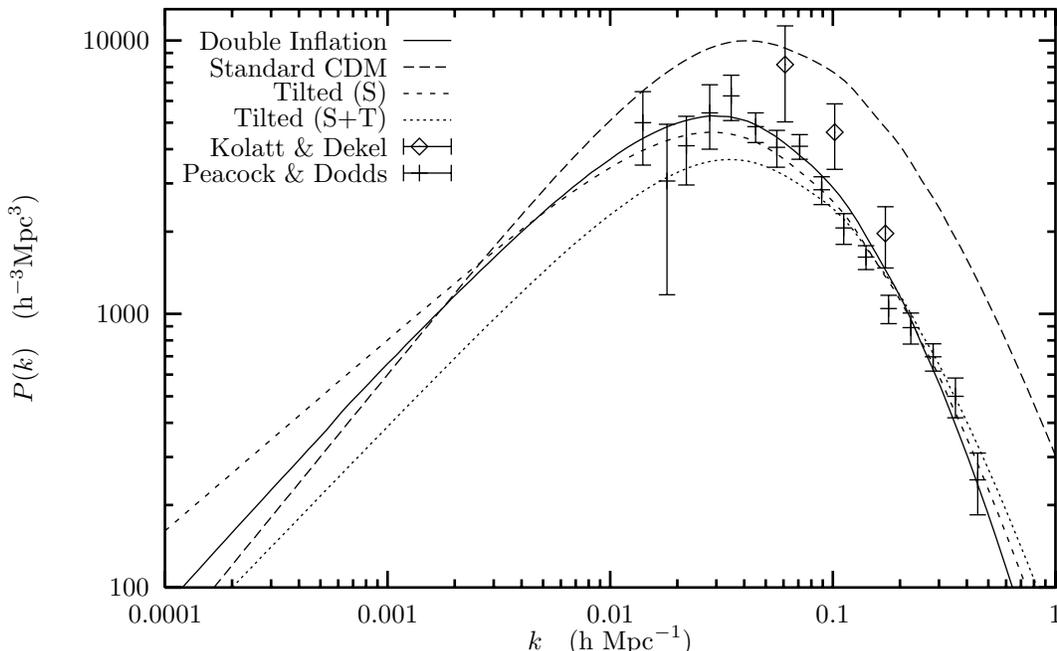

The power spectra corresponding to these four models are plotted in
figure \ref{fig.pow.spec}. 
In order to visualize the expected {\it shape} of the spectrum,
we also plot the redshift surveys data compilation of 
Peacock and Dodds \cite{PD}, rescaled to $\sigma_8=0.60$.
This plot clearly shows that the sucess of our double-inflationary
model in this respect is linked to the decrease of the effective index 
towards growing $k$, when $k<2\,h\,{\rm Mpc}^{-1}$ (see eq.(\ref{eq:log})).

\section{Cosmic Microwave Background anisotropies}\label{sec.4}

We compute the scalar and tensor CMB anisotropies using 
{\sc cmbfast}, the fast boltzmann code by  
Seljak and Zaldarriaga \cite{SZ}, extended to non-power-law
primordial spectra. 

The temperature anisotropies multipoles in Fourier space are calculated
at the present time, for scalar and tensor components: 
$\Delta_{Tl}^{(S,T)}(k)$. Taking into account the correct normalization
factors, and Fourier conventions, the scalar and tensor
modes multipoles then read in our notation:

\begin{equation}
C_l^S=\frac{2}{\pi} \int\!\! dk\,k^2 \phi^2(k)|\Delta_{Tl}^{(S)}(k)|^2~,
\qquad
C_l^T=\frac{1}{2\pi} \int\!\! dk\,k^2 h^2(k)|\Delta_{Tl}^{(T)}(k)|^2~.
\end{equation}

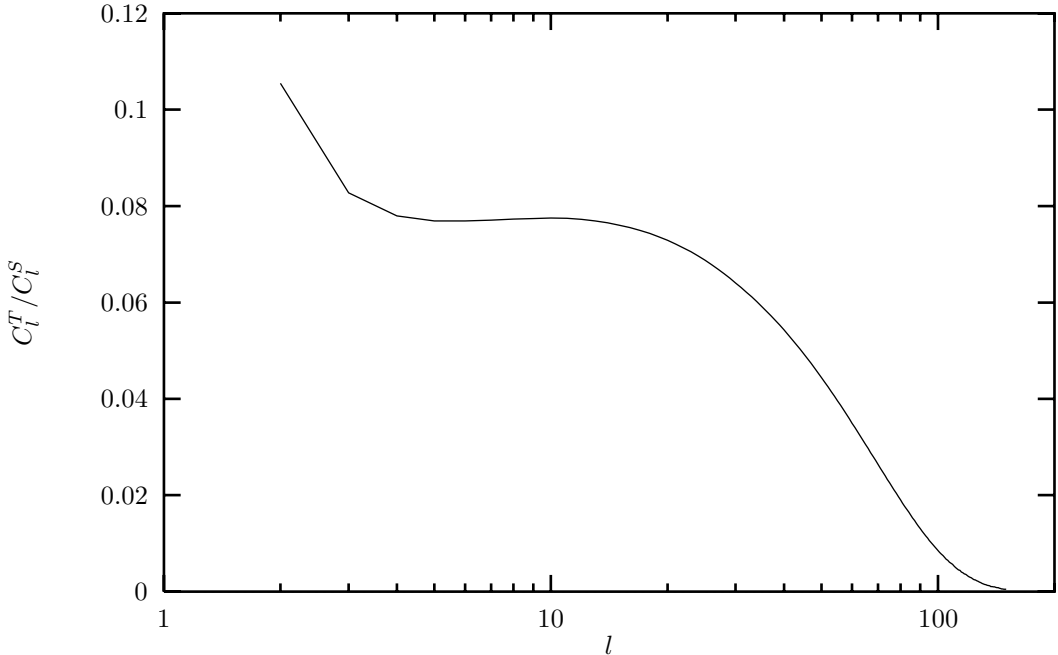
\begin{figure}[htbp]
\setlength{\unitlength}{0.1bp}
\begin{picture}(4139,2483)(0,0)
\put(2278,51){\makebox(0,0){$l$}}
\put(100,1341){%
\makebox(0,0)[b]{\shortstack{$C_l^T/C_l^S$}}%
}
\put(3517,151){\makebox(0,0){100}}
\put(2058,151){\makebox(0,0){10}}
\put(600,151){\makebox(0,0){1}}
\put(540,2432){\makebox(0,0)[r]{0.12}}
\put(540,2069){\makebox(0,0)[r]{0.1}}
\put(540,1705){\makebox(0,0)[r]{0.08}}
\put(540,1341){\makebox(0,0)[r]{0.06}}
\put(540,978){\makebox(0,0)[r]{0.04}}
\put(540,615){\makebox(0,0)[r]{0.02}}
\put(540,251){\makebox(0,0)[r]{0}}
\end{picture}
\caption[]{\small 
Ratio of tensor and scalar multipoles $C_l^T/C_l^S$ for double inflation. We 
have, in particular, $C_2^T/C_2^S=0.1\ll 7~|n_T|\simeq 7~(1-n)\approx 1$ in 
contrast with single-field slow-roll inflation.} 
\label{fig.ratio}
\end{figure}

We first plot in figure \ref{fig.ratio} the ratio of tensor and scalar
multipoles for parameters inside the allowed window. 

We see that $C_{10}^T/C_{10}^S=0.077$, in excellent agreement with
the analytic result of \cite{PLB95}. As already claimed in \cite{PLB95},
the relation $C_{2}^T/C_{2}^S\approx 7(1-n)$ is strongly 
violated in our case for which $C_{2}^T/C_{2}^S=0.1$\ while $n_{eff}\simeq 
0.85$ for large scales.

We then compute the total anisotropies for various parameters
inside the allowed window. The anisotropies only depend on 
$p$\ and $k_b$\ through the primordial power spectrum, so they increase with $k_b$\ and hardly depend on $p$ at relevant scales. 
In fact they increase very slightly with $p$\ since 
$2 k_b / a_0 H_0 \gg 1500$ and hence all the l's containing the three peaks 
correspond to scales, much smaller than $k_b$, where the primordial spectra 
do not differ much as can be seen in figure 1. Since on the other hand there 
is a precise constraint on $k_b$, we obtain sharp predictions for the 
multipoles. For instance, the position and amplitude of the first two peaks 
are within the ranges:
\begin{enumerate}
\item
for the first peak, located at $l=207 \pm 1$,
\begin{equation} 
\frac{l(l+1)C_l}{110C_{10}}
= 2.6 \pm 0.1~;
\end{equation}
\item
for the second peak, located at $l=505$,
\begin{equation}
\frac{l(l+1)C_l}{110C_{10}}
= 1.2 \pm 0.1~.
\end{equation}
\end{enumerate}

\begin{figure}[htbp]
\setlength{\unitlength}{0.1bp}
\begin{picture}(4139,2483)(0,0)
\put(1274,1658){\makebox(0,0)[r]{CAT}}
\put(1274,1758){\makebox(0,0)[r]{MAX}}
\put(1274,1858){\makebox(0,0)[r]{ARGO}}
\put(1274,1958){\makebox(0,0)[r]{Saskatoon}}
\put(1274,2058){\makebox(0,0)[r]{South Pole}}
\put(1274,2158){\makebox(0,0)[r]{Tenerife}}
\put(1274,2258){\makebox(0,0)[r]{COBE}}
\put(2278,51){\makebox(0,0){$l$}}
\put(100,1341){%
\makebox(0,0)[b]{\shortstack{$\left( l(l+1)C_l\,/\,2 \pi \right)^{1/2}$}}%
}
\put(3619,151){\makebox(0,0){1000}}
\put(2501,151){\makebox(0,0){100}}
\put(1382,151){\makebox(0,0){10}}
\put(540,2432){\makebox(0,0)[r]{3e-05}}
\put(540,1996){\makebox(0,0)[r]{2.5e-05}}
\put(540,1560){\makebox(0,0)[r]{2e-05}}
\put(540,1123){\makebox(0,0)[r]{1.5e-05}}
\put(540,687){\makebox(0,0)[r]{1e-05}}
\put(540,251){\makebox(0,0)[r]{5e-06}}
\end{picture}
\caption[]{\small 
CMB anisotropies are plotted for several models (from top to bottom at 
$l$=200): 
standard CDM, double-inflationary model ($p=12,~k_b=2.3~h~{\rm Mpc}^{-1}$), 
tilted model without tensors ($n=0.70$),
tilted model with tensors ($n=0.845$). Error bars are at the $1\sigma$ level. 
The measurements of the following experiments are indicated, in order of 
appearance for growing $l$: COBE (3 points), Tenerife, South Pole, 
Saskatoon (5 points), MAX (2 points), MSAM, CAT (2 points), and OVRO. 
For the double-inflationary model, we have 
in particular $l(l+1)C_l\approx 2.6\times 110 C_{10}$ for $l\approx 200$.} 
\label{fig.cl}
\end{figure}
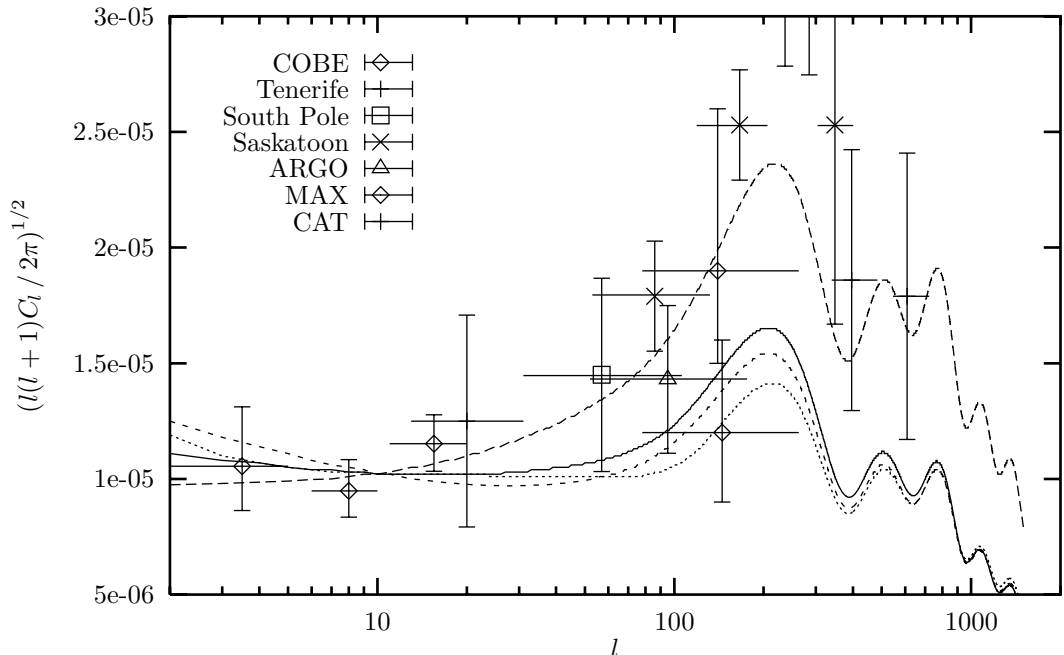

The full $(l(l+1)C_l/2 \pi)^{1/2}$\ curve is given 
on figure \ref{fig.cl} for
$p=12$\ and $k_b=2.3\,h\,Mpc^{-1}$,
together with standard CDM and tilted models.
 
We include a few measurements of the anisotropies:
COBE \cite{HI}, Tenerife \cite{GU2},  
South Pole \cite{GU}, 
Saskatoon (with the recent recalibration) \cite{NE,LE1},
MAX \cite{TA}, MSAM's third flight \cite{CH}, 
and new preliminary points from
CAT \cite{BA} and OVRO \cite{LE2}.
A complete analysis of the presently available data set can be found in
\cite{LB}. 

As expected from the previous section, the amplitude of the peaks
is higher in the double-inflationary case than in the tilted one.
However, it is clear from the figure that the improvement
is not sufficient to agree with observations at little scales.
The Doppler peak is approximately at
$4.5\sigma$\ under Saskatoon and $2\sigma$\ under MSAM measurements. The 
secondary acoustic peaks, or Sakharov oscillations, are around
$2\sigma$\ under CAT and Ovro.
 
In the case of a higher multipole normalization, already considered
in Sec. \ref{sec.pow.spec}, we obtain a slight increase of the first
and third peaks, however not a very significant one, since the
global increase is damped by the shift of allowed $k_b$'s to lower values.
It is clear then that our 
model is not viable in the context of CDM with standard values of the 
cosmological parameters. 

Let us see wether this remains true when small variations of $h$\ and 
$\Omega_b$ are considered. Keeping $h$\ fixed, an increase in $\Omega_b$\
will enhance the radiation transfer function on the 
one hand and damp the matter transfer function on the other hand, 
therefore requiring higher $k_b$'s in order to keep
enough power and satisfy large scale strucure tests. Both effects 
contribute to increase CMB anisotropies. If $\Omega_B h^2=0.025$,
within the corresponding allowed window,
the highest peak is given by $p=12$, $k_b=4\,h\,\,\mathrm{Mpc}^{-1}$:
\begin{equation} 
\frac{l(l+1)C_l}{110C_{10}}
= 3.5, \qquad l=213
\end{equation}
This improvement is too small to agree with the observations:
the peak is still $3\sigma$\ under Saskatoon and 
more than $1\sigma$ under MSAM.

Going to higher $h$, with $\Omega_B h^2$\ fixed,
just yields the opposite effect: the radiation transfer function
and the shift in the allowed parameters both lower the anisotropies.
When $h=0.6$\ and $\Omega_B h^2=0.015$\ (resp. $0.025$), the peak is as low
as $l(l+1)C_l / 110C_{10}= 1.9$\ (resp. $2.5$) in the best case.   

\section{Conclusion}\label{sec.5}

We have studied here a model of double inflation using constraints from 
both large scale surveys and CMB anisotropies. The primordial spectrum in 
such a model has a characteristic length with more power towards large scales.
The primary purpose of this model is to reconcile CDM with observations in 
the following sense: we keep the canonical values of standard CDM, possibly 
allowing a small departure from these, and we take 
the primordial spectrum of scalar perturbations generated during the 
inflationary stage in our model instead of a nearly scale invariant (Harrison
Zel'dovich) spectrum. So, it should be stressed that once the underlying 
theory is given, there are two more free parameters as compared to standard 
CDM, hence all the spectra, including that one of the gravitational waves, are 
computed from first principles, and no ad hoc assumptions are made. 

The observations of the CMB anisotropies on the one hand and of the density 
fluctuation power spectrum $P(k)$ on the other hand, give independent tests 
of the primordial fluctuations spectra.
It turns out that the power spectrum $P(k)$ in our model meets well the large 
scales structures observations, as summarized in figure \ref{fig.domain}, 
though as found 
earlier the peculiar velocities are low. It does certainly better 
than tilted models if one insists on the value $\sigma_8\approx 0.6 $.     
We have also studied the CMB anisotropies produced in our model and we find 
that the observations of CMB anisotropies on intermediate and small angular 
scales tremendously constraint our model. The Saskatoon and
MSAM data on angular 
scales corresponding to the Doppler peak seem to imply a much higher 
peak than obtained in our model. If these observations are to be confirmed, 
then it is clear that in the framework of a flat CDM universe with values 
of the cosmological parameters close to those of standard CDM, our model is 
to be excluded. Note that the computed 
CMB anisotropies are obtained using linear perturbation theory so that the  
results are very reliable and put severe constraints on any model if the 
CMB anisotropies on those angular scales are measured with great accuracy 
even after cosmic variance is taken into account. This is the goal of 
the future satellite experiments {\it PLANCK Surveyor} and {\it MAP} and their 
observations will tremendously constrain all models proposed. 
As stressed above, the observations of CMB anisotropies on the one hand and of 
the density fluctuation power spectrum, on the other hand, give independent 
constraints of the primordial fluctuations spectra.

One could then 
ask whether this implies that all inflationary models yielding a primordial 
spectrum with a characteristic scale should be rejected, again if we keep 
the same CDM (flat) universes. Actually, this turns out not to be the case. 
Indeed, the problems in our model arise due to the form of the spectrum on 
large scales. Therefore while all models with a spectrum analogous 
to ours will yield similar CMB anisotropies \cite{AA93} and are therefore 
to be excluded, on the other hand, a spectrum closer to a steplike 
spectrum with a flat or slightly increasing $(n>1)$ upper plateau is expected 
to give better agreement with observations. A possibility, based on 
the spatial distribution of rich Abell clusters, is a spike in the initial 
spectrum followed by a stepdown as proposed in \cite{nat97}.

\end{document}